\documentclass[onecolumn]{aa}
\usepackage{graphicx}
\usepackage{txfonts}
\begin{document}

   \title{The progenitors of type Ia supernovae in the semidetached binaries with red giant donors}


   \author{D. Liu\inst{1,2,3,4},
          B. Wang\inst{1,2,3,4},
          H. Ge\inst{1,2,3,4},
          X. Chen\inst{1,2,3,4}
          \and
          Z. Han\inst{1,2,3,4}
          }
   \institute{Yunnan Observatories, Chinese Academy of Sciences, Kunming 650216, China\\
              \email{liudongdong@ynao.ac.cn; wangbo@ynao.ac.cn}
          \and
          Key Laboratory for the Structure and Evolution of Celestial Objects, Chinese Academy of Sciences, Kunming 650216, China
          \and
          University of Chinese Academy of Sciences, Beijing 100049, China
          \and
          Center for Astronomical Mega-Science, Chinese Academy of Sciences, Beijing, 100012, China}

   \date{}

  \abstract
   {The companions of the exploding carbon-oxygen white dwarfs (CO WDs) for producing type Ia supernovae (SNe Ia) are still not conclusively confirmed. A red-giant (RG) star has been suggested to be the mass donor of the exploding WD, named as the symbiotic channel. However, previous studies on the this channel gave a relatively low rate of SNe Ia.}
   {We aim to systematically investigate the parameter space, Galactic rates and delay time distributions of SNe Ia from the symbiotic channel by employing a revised mass-transfer prescription.}
   {We adopted an integrated mass-transfer prescription to calculate the mass-transfer process from a RG star onto the WD. In this prescription, the mass-transfer rate varies with the local material states.}
   {We evolved a large number of WD$+$RG systems, and found that the parameter space of WD$+$RG systems for producing SNe Ia is significantly enlarged. This channel could produce SNe Ia with intermediate and old ages, contributing to at most 5\% of all SNe Ia in the Galaxy. Our model increases the SN Ia rate from this channel by a factor of 5. We suggest that the symbiotic systems RS Oph and T CrB are strong candidates for the progenitors of SNe Ia.}

   {}

\titlerunning{The progenitors of SNe Ia in the semidetached binaries with RG donors}

\authorrunning{D. Liu et al.}

   \maketitle
%

\section{Introduction} \label{1. Introduction}
Type Ia supernovae (SNe Ia) are good distance indicators for cosmology, which reveals the accelerating expansion of the Universe and leads to the discovery of dark energy (e.g. Howell 2011; Meng et al. 2015). It is generally believed that SNe Ia result from thermonuclear explosions of carbon-oxygen white dwarfs (CO WDs) in binaries (Hoyle \& Fowler 1960). However, the mass donor for the exploding CO WD is still not fully confirmed (e.g., Podsiadlowski et al. 2008; Howell 2011; Wang \& Han 2012; Maoz et al. 2014). The mass donor could be a non-degenerate star in the single-degenerate model (e.g., Schatzman 1963; Truran \& Cameron 1971; Whelan \& Iben 1973), or another WD in the double-degenerate model (e.g., Iben \& Tutukov 1984; Webbink 1984; Liu et al. 2016, 2017, 2018).

In the single-degenerate model, the primary WDs can accrete H-rich matter from red-giant (RG) stars and form SNe Ia when the WDs grow in mass close to the Chandrasekhar limit ($M_{\rm Ch}$), known as the symbiotic channel (e.g., Whelan \& Iben 1973; Kenyon et al. 1986, 1993; Munari \& Renzini 1992; Yungelson \& Livio 1998; King et al. 2003). Although the real number of symbiotic stars in the Galaxy is still unknown (e.g., L\"u et al. 2006; Miko\l{}ajewska 2012; Rodr\'{i}guez-Flores et al. 2014), there are still many symbiotic systems in the observations (e.g., Belczy\'nski et al. 2000; Miszalski \& Miko\l{}ajewska 2014; Li et al. 2015), in which the symbiotic systems T CrB (Kraft 1958) and  RS Oph (Brandi et al. 2009) are possible progenitor candidates for SNe Ia. Patat et al. (2007) detected Na I absorption lines with low expansion velocities in SN 2006X, and speculated that the companion of the exploding WD may be an early RG star, although Chugai (2008) argued that the absorption lines detected in SN 2006X cannot be formed in the RG wind.
Voss \& Nelemans (2008) suggested that the progenitor of SN 2007on may be a WD$+$RG system after studying the pre-explosion X-ray images at the same position.
In addition, the surviving companions of SNe Ia from the symbiotic channel may be related to the formation of single low-mass He WDs in the observations (e.g., Justham et al. 2009; Wang \& Han 2010).

However, previous studies argued that the rate of SNe Ia from the symbiotic channel is relatively low (Li \& van den Heuvel 1997; Yungelson \& Livio 1998; Han \& Podsiadlowski 2004). These studies usually assumed that the exceeding mass of the RG star would be transferred onto the surface of the WD rapidly once the RG star exceeds its Roche-lobe (e.g., Han et al. 2000). In this case, the mass-transfer rate ($\dot{M}_{\rm 2}$) is usually relatively high, resulting in two cases that prevent forming SNe Ia. (1) A common envelope (CE) may be formed if the mass-transfer is dynamically unstable. After the CE ejection, the binary may evolve to a CO WD$+$He WD system. (2) The binary may enters a stellar wind stage, in which too much matter are blown away from the system that would prevent the WD from growing in mass to $M_{\rm Ch}$.

We note that the local gas density and sound velocity in the shell of a RG star are significantly lower than that in a main-sequence (MS) star. However, previous studies usually employed the same mass-transfer prescription for WD$+$RG systems, which may overestimate $\dot{M}_{\rm 2}$ when the RG star fills its Roche-lobe.
Lubow \& Shu (1975) proposed that the mass-transfer process can be investigated by integrating the local gas density and sound velocity over the plane that is perpendicular to the centre line connecting the two stars, and passing through the inner Lagrangian point. By assuming that the state equation of stars obeys a power-law adiabat and the mass outflow is laminar and occurs along equipotential surface, Ge et al. (2010) obtained an approximation prescription for the mass-transfer process, which is similar to that presented in Kolb \& Ritter (1990). In this prescription, $\dot{M}_{\rm 2}$ is relevant to the local material states, and the exceeding mass will be not transferred immediately when the RG star fills its Roche-lobe. $\dot{M}_{\rm 2}$ is lower than that of previous models, leading to the results that the mass loss turns to be lower, and that more matter are accumulated onto the primary WD.
Ge et al. (2015) also found that the critical mass ratio for the formation of a CE becomes larger, which reveals that more interacting binaries would experience stable mass-transfer process.

In this Letter, we adopted an integrated mass-transfer prescription described in the Appendix of Ge et al. (2010) to investigate the semidetached CO WD$+$RG channel for producing SNe Ia. In Sect.\,2, we show the methods for binary evolution computations and the corresponding results. The methods and results for binary population synthesis (BPS) are provided in Sect.\,3. We present a discussion and summary in Sect.\,4.

\section{Binary evolution computations} \label{2. Binary evolution calculations}
\subsection{Methods}
We use the Eggleton stellar evolution code (Eggleton 1973) to trace the binary evolutions of semidetached WD$+$RG systems. The typical Pop I composition is adopted for the initial MS models with H fraction $X=0.7$, He fraction $Y=0.28$ and metallicity $Z=0.02$. When the WD grows in mass to $M_{\rm Ch}$ (set to be $1.378\, M_{\odot}$), an SN Ia explosion is assumed to occur.
In this work, we adopted the integrated mass-transfer prescription presented in the Appendix of Ge et al. (2010) to calculate the Roche-lobe overflow in WD$+$RG systems. The mass-transfer rate
\begin{equation}
\dot{M}_{\rm 2}=\frac{2\pi R_{\rm L}^{\rm 3}}{GM_{\rm 2}}f(q)
\int_{\rm \phi_{\rm L}}^{\rm \phi_{\rm s}}\Gamma^{\rm 1/2}(\frac{2}{\Gamma+1})^{\rm \frac{\Gamma+1}{2(\Gamma-1)}}(\rho P)^{\rm 1/2} \rm d\phi,
\end{equation}
where $R_{\rm L}$ is the effective Roche-lobe radius, G is the gravitational constant, $M_{\rm 2}$ is the donor mass, $\Gamma$ is the adiabatic index, $\rho$ is the local gas density and $P$ is the local gas pressure. The integration is from the Roche-lobe potential energy ($\phi_{\rm L}$) to the stellar surface potential energy ($\phi_{\rm s}$).
The coefficient $f(q)$ is a slowly-varying function of the mass ratio $q$:
\begin{equation}
f(q)\equiv\frac{q}{r_{\rm L}^{\rm 3}(1+q)}\frac{1}{[a_{\rm 2}(a_{\rm 2}-1)]^{\rm 1/2}},
\end{equation}
in which $a_{\rm 2}$ is a function of $q$ (see the Equations A3 and A4 in Ge et al. 2010).
The potential energy $\rm d\phi=GM_{\rm 2}R^{\rm -2}\rm dR$, in which $R$ is the donor radius.

The accreted H-rich matter from RG stars first burns into He. The formed He experiences He flash and burns into C and O accumulating onto the surface of the WD. The mass growth rate of the WDs ($\dot {M}_{\rm WD}$) is defined as
\begin{equation}
\dot{M}_{\rm WD}=\eta_{\rm He}\eta_{\rm H}\dot{M}_{\rm 2},
\end{equation}
where $\eta_{\rm H}$ is the mass accumulation efficiency for H-shell burning from Wang et al. (2010), and $\eta_{\rm He}$ is the mass accumulation efficiency for He-shell flashes from Kato \& Hachisu (2004). The value of $\eta_{\rm He}$ may be overestimated by this prescription when the Roche-lobe overflow is taken into account (e.g., Piersanti et al. 2014). The same effect may be for $\eta_{\rm H}$, which needs further exploration. When $\dot{M}_{\rm 2}$ is larger than the critical mass-transfer rate $\dot {M}_{\rm cr}$ described in Nomoto (1982), we assume that the mass-growth rate of the WD is $\dot {M}_{\rm cr}$ and the rest of H-rich matter would be blown away in the form of the optically thick wind (e.g., Hachisu et al. 1996). We also assume that the mass loss takes away the specific orbital angular momentum of the primary WD.

\subsection{Results}
\begin{figure*}
   \centering
   \includegraphics[width=7.5cm,angle=270]{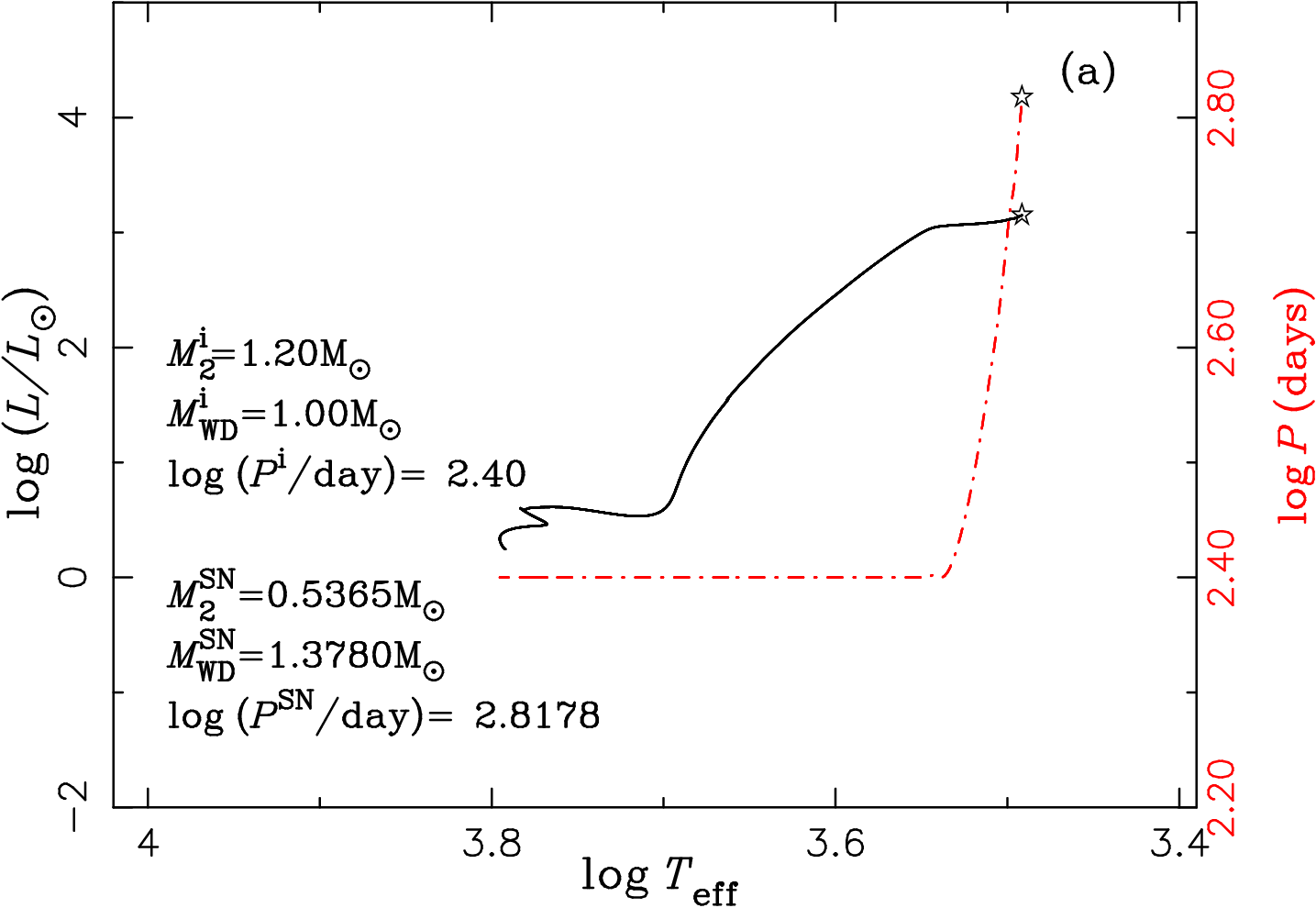}
   \includegraphics[width=7.5cm,angle=270]{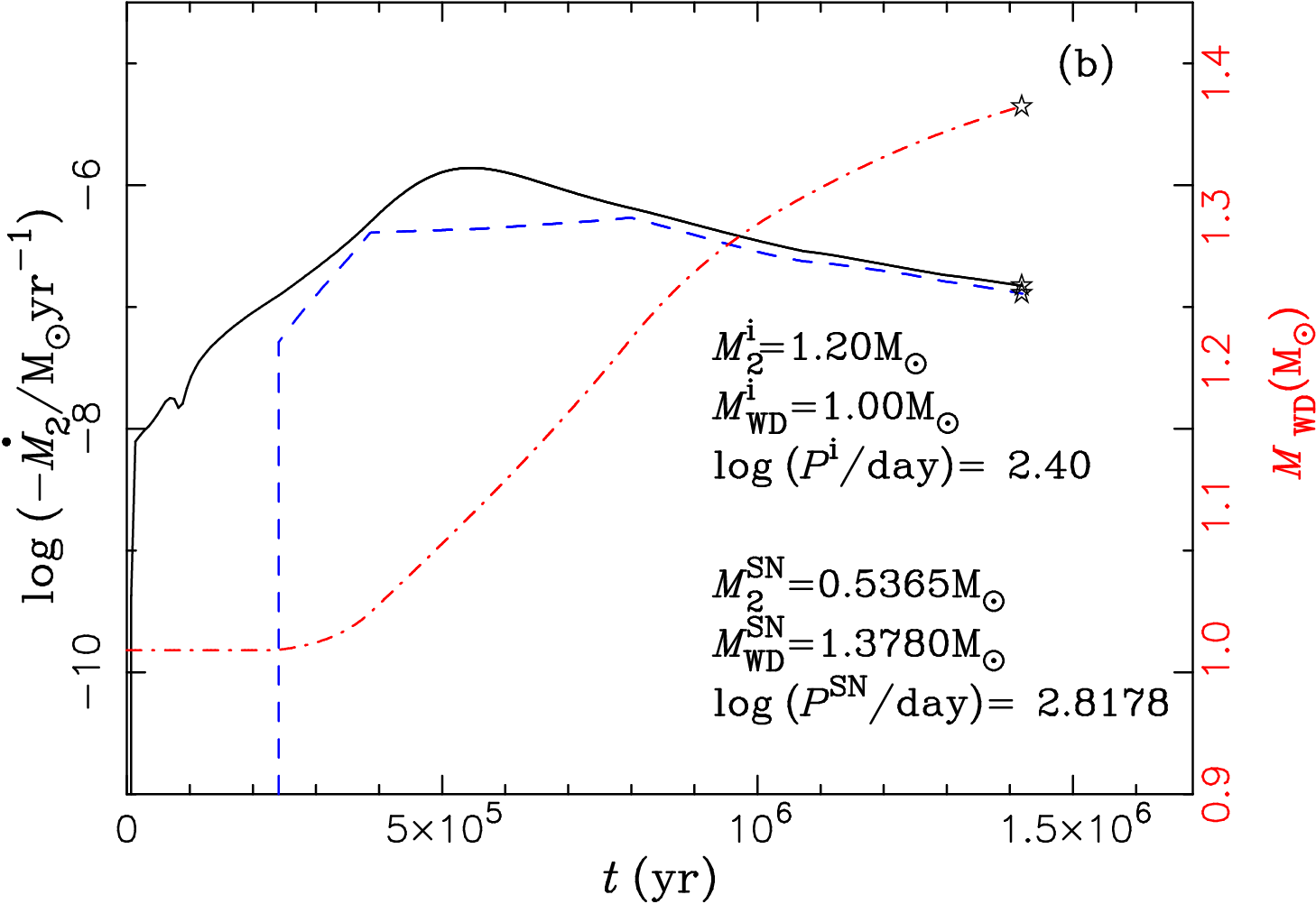}
   \caption{A typical binary evolution for producing SN Ia based on the symbiotic channel.
  In panel (a), the solid curve represents the evolutionary track of the mass donor in the Hertzsprung-Russell diagram, and the dash-dotted curve shows the evolution of the orbital periods.
  In panel (b), the evolution of $\dot{M}_{\rm 2}$, $\dot{M}_{\rm WD}$ and $M_{\rm WD}$ as a function of time are shown as solid, dashed and dash-dotted curves, respectively.
  The asterisks in both panels indicate the position where SN Ia explosion occurs.}
\end{figure*}

Fig.\,1 shows a typical example of binary evolution computations, in which the WD accretes H-rich material from a RG star and explodes as an SN Ia when its mass reaches $M_{\rm Ch}$. Here, we adopt the initial WD mass $M_{\rm WD}^{\rm i}=1.0\, M_{\odot}$, the initial mass of donor star $M_{\rm 2}^{\rm i}=1.2\, M_{\odot}$ and the orbital period $\log\,(P^{\rm i}/\rm day)=2.4$. When the mass donor evolves to the RG stage, it expands quickly and fills its Roche-lobe. At the beginning, $\dot{M}_{\rm 2}$ is relatively low and no matter is accumulated due to strong H-shell flashes. After about $2.3\times10^{\rm 5}\,\rm yr$, the binary enters the optical thick wind stage, in which the transferred H-rich matter burns into C and O and accumulates onto the primary WD at the rate of $\dot{M}_{\rm cr}$, while the rest of H-rich matter is assumed to be blown away via the optically thick wind. After about $4.1\times10^{\rm 5}\,\rm yr$, the binary enters the stable H burning stage. The primary WD grows in mass to $M_{\rm Ch}$ and explodes as an SN Ia about $6.2\times10^{\rm 5}\,\rm yr$ later. At this moment, the mass of RG star is $M_{\rm 2}^{\rm SN}=0.5365\, M_{\odot}$ with a $0.42\, M_{\odot}$ He core, and the orbital period is $\log\,(P^{\rm SN}/\rm day)=2.8178$.

\begin{figure}
   \centering
   \includegraphics[width=7.5cm,angle=270]{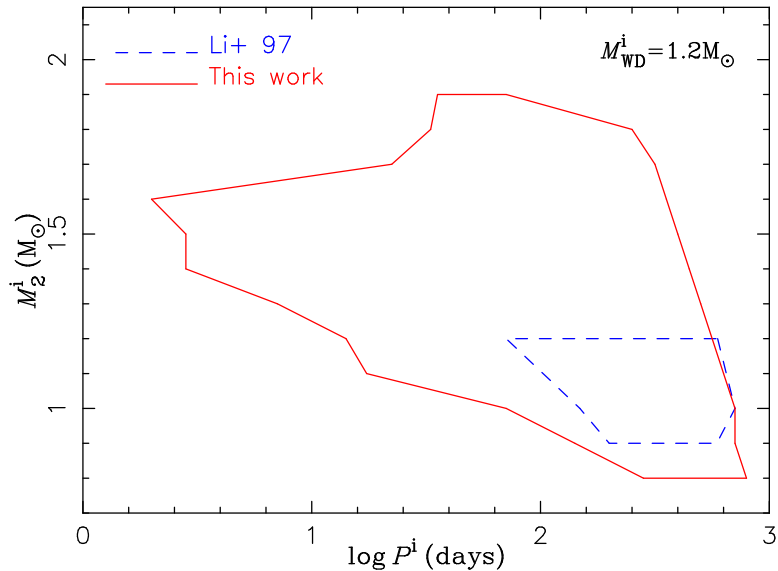}
 \caption{Regions of WD$+$RG systems for producing SNe Ia with $M_{\rm WD}^{\rm i}=1.2\, M_{\odot}$. The red solid contour shows the parameter space from this work, whereas the blue dashed contour represents that from Li \& van den Heuvel (1997) for a comparison.}
\end{figure}

We evolved a large sample of WD$+$RG systems, and obtained the initial parameter space for the production of SNe Ia via the symbiotic channel. Fig.\,2 shows the regions of WD$+$RG systems in the initial orbital period$-$initial secondary mass ($\log P^{\rm i}-M^{\rm i}_{\rm 2}$) plane for producing SNe Ia with an initial WD mass of $1.2\, M_{\odot}$. For a comparison, we show the results of Li \& van den Heuvel (1997). This figure shows that the grid from the present work have larger initial MS masses and shorter initial orbital periods. In these regions, $\dot{M}_{\rm 2}$ would be so high that the binary enters a CE process or strong optically thick wind process in the model of Li \& van den Heuvel (1997), preventing to form SNe Ia.

\begin{figure}
   \centering
   \includegraphics[width=7.5cm,angle=270]{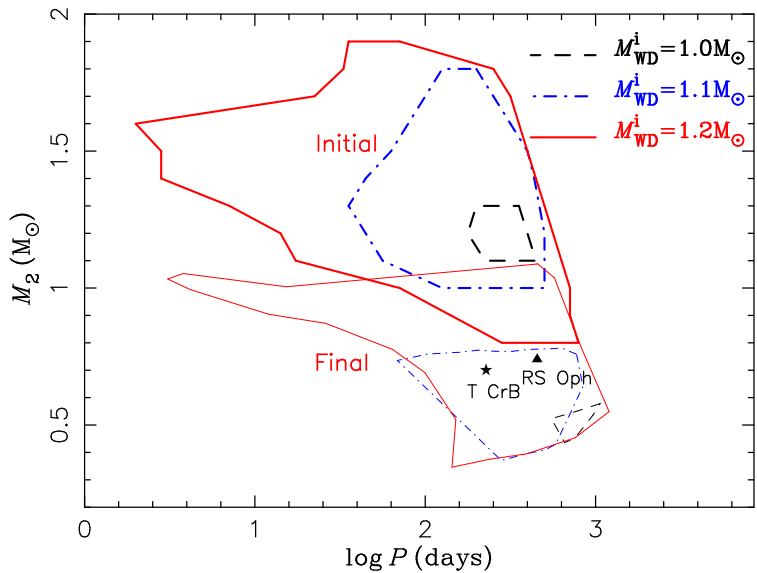}
 \caption{Initial and final regions of WD$+$RG systems for producing SNe Ia with various $M_{\rm WD}^{\rm i}$. The filled triangle and asterisk represent two symbiotic systems RS Oph and T CrB, respectively.}
\end{figure}

Fig.\,3 presents the initial parameter space for the production of SNe Ia, and the final regions of WD+RG systems at the moment of SN Ia explosions in the orbital period-secondary mass ($\log P-M_{\rm 2}$) plane, in which $M_{\rm WD}^{\rm i}=1.0, 1.1$ and $1.2\, M_{\odot}$. We found that $M_{\rm WD}^{\rm i}=1.0\, M_{\odot}$ is the minimum initial WD mass for producing SNe Ia as its region almost vanishes.
We also found that the symbiotic systems RS Oph and T CrB could form SNe Ia via the symbiotic channel (for more details see Sect.\,4).
The WDs in the binaries beyond these initial contours can not grow in mass to $M_{\rm Ch}$: binaries beyond the left boundaries will experience strong H-shell flashes that will blow away too much material, while binaries beyond the right boundaries will undergo rapid mass-transfer process as a result of the rapid expansion of the RG stars, leading to the mass-loss of too much material in the form of the optically thick wind. Note that the mass donor with the shortest orbital period for the case of $M_{\rm WD}^{\rm i}=1.2\, M_{\odot}$ would fill its Roche-lobe on the bottom of its RG branch. The lower boundaries are determined by the conditions that $M_{\rm 2}$ and $\dot{M}_{\rm 2}$ should be larger, so that the primary WDs can grow in mass to $M_{\rm Ch}$. The upper boundaries are constrained by the high mass-transfer rate owing to the large mass-ratio, which may lead to the formation of a CE. The wind-accreting channel of symbiotic stars may also slightly contribute the formation of SNe Ia, which is not considered in this work as its contribution is almost negligible (e.g., Yungelson et al. 1995).

\section{Binary population synthesis} \label{3. BPS}
\subsection{Methods}
By employing the Hurley rapid binary evolution code (Hurley et al. 2002), we conduct a series of Monte Carlo simulations in the BPS approach to calculate the rates and delay times of SNe Ia. In each of our simulation, we investigate the evolution of $4\times10^7$ primordial samples until the formation of WD$+$RG systems. The metallicity in our simulations is set to be 0.02. We assume that an SN Ia will be produced if the parameters of the formed WD$+$RG system are located in the initial regions for producing SNe Ia in Fig.\,3.

The initial parameters and basic assumptions of the Monte Carlo BPS computations listed below are adopted. (1) All stars are assumed to be in binary systems with circular orbits. (2) The initial mass function of Kroupa (2001) is adopted for the primordial primaries. (3) The primordial mass ratio distribution is assumed to be constant. (4) The initial distribution of separations $a$ is supposed to be constant in log a for wide binaries, and fall off smoothly for close binaries (see Eggleton et al. 1989).\footnote{Note that there is still no serious work with account of selection effects on the distributions of the primordial mass ratio and orbital separation since the late 1980s.} (5) The standard energy prescription from Webbink (1984) is employed to describe the CE ejection process, in which the uncertain parameters $\alpha_{\rm CE}$ and $\lambda$ are combined as a single parameter and set to be $\alpha_{\rm CE}\lambda=0.5, 1.0$ and 1.5 for comparison. (6) The star formation rate is adopted to be constant ($5\, M_{\odot}yr^{\rm -1}$) for the Galaxy  over the past $15\,\rm Gyr$ (see Yungelson \& Livio 1998; Willems \& Kolb 2004; Han \& Podsiadlowski 2004), or alternatively, modeled as a delta function (a single star burst of $10^{\rm 10}\, M_{\odot}$ in stars). Note that Chomiuk \& Povich (2011) suggested that the current Galactic star formation rate is about $1.9\pm0.4\, M_{\odot}yr^{\rm -1}$. Thus, we also employ a constant star formation rate of $2\, M_{\odot}yr^{\rm -1}$ for comparison.

The formation channel of semidetached WD$+$RG systems has been investigated by Tutukov \& Yungelson (1976), Kenyon \& Webbink (1984), Yungelson et al. (1995), Wang et al. (2010), etc. The primordial primary fills its Roche-lobe when it evolves to the thermal pulsing asymptotic giant branch. In this case, the mass-transfer is dynamically unstable, leading to the formation of a CE. If the CE can be ejected, the primordial primary becomes a CO WD. After that, a WD$+$RG system will be formed when the primordial secondary evolves to its RG phase. The parameters of primordial binaries are $M_{\rm 1,i}$$\sim$5.0$-$$6.5\, M_{\odot}$, $0.15<M_{\rm2,i}/M_{\rm1,i}<0.5,$ and $P^{\rm i}$$\sim$600$-$$5000\,\rm days$ for producing SNe Ia via the symbiotic channel.

\subsection{Results}
\begin{figure}
   \centering
   \includegraphics[width=7.5cm,angle=270]{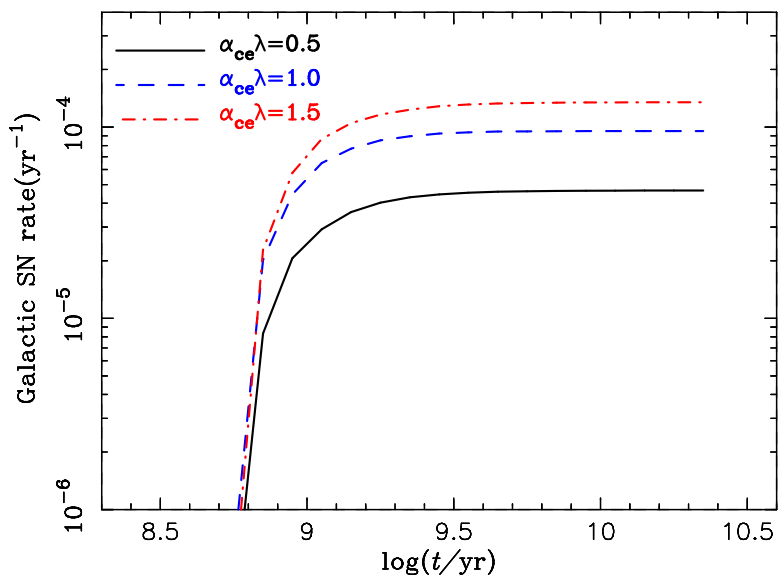}
 \caption{Evolution of SN Ia rates in the Galaxy as a function of time based on the symbiotic channel. The solid, dashed and dash-dotted curves represent the cases with $\alpha_{\rm CE}\lambda=0.5$, 1.0 and 1.5, respectively.}
\end{figure}

Fig.\,4 presents the evolution of SN Ia rates from the symbiotic channel with different values of $\alpha_{\rm CE}\lambda$. The star formation rate here is adopted to be $5\, M_{\odot}yr^{\rm -1}$.
This figure shows that the Galactic rates of SNe Ia are in the range of $\sim$0.5$-$$1.3\times10^{\rm -4}\,\rm yr^{\rm -1}$, contributing to at most 5\% of all SNe Ia in the Galaxy. Note that the SN Ia rate increases with $\alpha_{\rm CE}\lambda$. This is because the orbital period of the binaries evolving from CE ejections would be larger for a longer $\alpha_{\rm CE}\lambda$, resulting in more WD$+$RG systems locating in the SN Ia production region in Fig.\,3.
If an identical star formation rate is adopted (i.e. $5\, M_{\odot}yr^{\rm -1}$), the traditional SN Ia rate from the symbiotic channel may contribute to about $1\%$ of all SNe Ia (e.g., Han \& Podsiadlowski 2004; Wang et al. 2010).
Obviously, our model increases the rate of SNe Ia from the symbiotic channel by a factor of 5.
If the Galactic star formation rate is assumed to be $2\, M_{\odot}$ (e.g., Chomiuk \& Povich 2011), the Galactic rates of SNe Ia from the symbiotic channel decrease to be $\sim$0.2$-$$0.6\times10^{\rm -4}\,\rm yr^{\rm -1}$, contributing to at most 2\% of all SNe Ia in the Galaxy.

\begin{figure}
   \centering
   \includegraphics[width=7.5cm,angle=270]{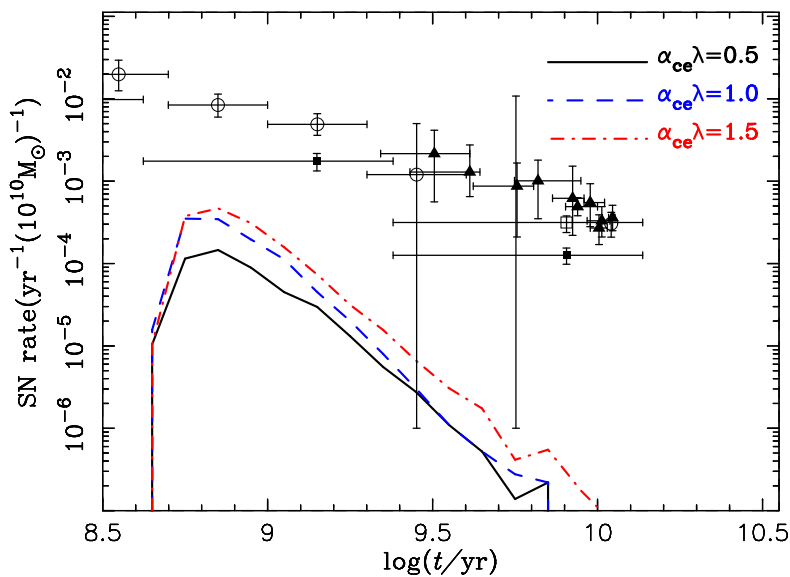}
 \caption{Delay time distributions of SNe Ia from the symbiotic channel, in which a star burst of $10^{\rm 10}\, M_{\odot}$ in stars is adopted. The open circles, the filled triangles, the filled squares, the open square represent observed results from Totani et al. (2008), Maoz et al. (2010, 2012) and Graur \& Maoz (2013), respectively.}
\end{figure}

Fig.\,5 shows the theoretical delay time distributions of SNe Ia from the symbiotic channel. A single starburst of $10^{\rm 10}\, M_{\odot}$ in stars is assumed here. According to the symbiotic channel, the SN Ia delay times range from about $400\,\rm Myr$ to $11\,\rm Gyr$, indicating that this channel mainly contributes to the observed SNe Ia in the intermediate and old populations. Note that Wang et al. (2010) suggested that the symbiotic channel only contributes to SNe Ia in old populations. The present work extends the contribution of the symbiotic channel to the SNe Ia in the intermediate populations.

\begin{figure}
   \centering
   \includegraphics[width=7.5cm,angle=270]{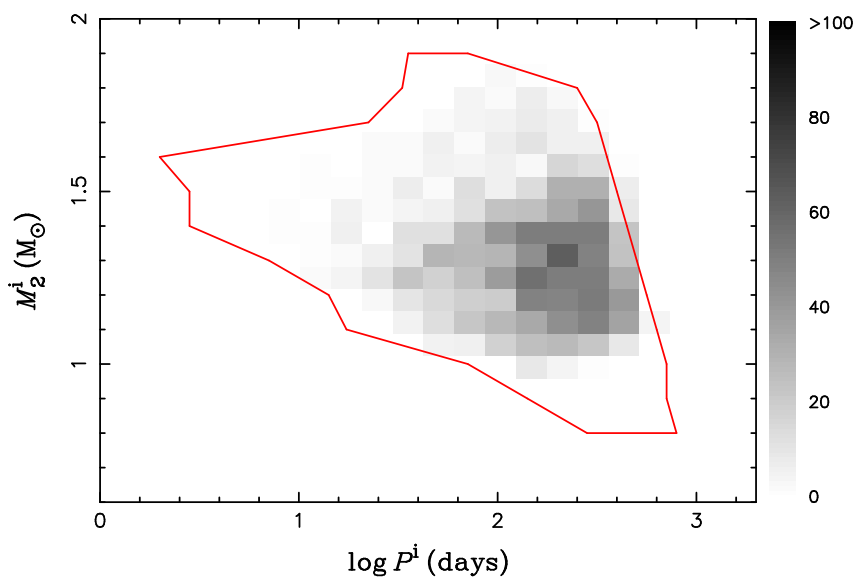}
 \caption{The density distribution of WD$+$RG systems for producing SNe Ia. Here, $\alpha_{\rm CE}\lambda$ is set to be 1.5 and $4\times10^{\rm 7}$ primordial samples are included. The solid contour is the initial parameter space of WD$+$RG systems for producing SNe Ia with $M^{\rm i}_{\rm WD}=1.2\, M_{\odot}$.}
\end{figure}

Fig.\,6 presents the density distribution of WD$+$RG systems for producing SNe Ia in the $\log P^{\rm i}-M^{\rm i}_{\rm 2}$ plane. For a comparison, we also show the initial contour for producing SNe Ia with $M^{\rm i}_{\rm WD}=1.2\, M_{\odot}$. From this figure, we can see that the distribution of the orbital periods of WD$+$RG systems mainly range from about $15\,\rm d$ to the right boundary of the initial contour, and the donor masses mainly distribute from $1.0\, M_{\odot}$ to the upper boundary of the initial contour. The contribution of the rest space to the SN Ia rates are negligible.

\section{Discussion and summary} \label{4. Discussion}
RS Oph and T CrB are two symbiotic systems that both consist a massive WD and a lobe-filling RG star. RS Oph has a $1.2-1.4\, M_{\odot}$ WD and a $0.68-0.8\, M_{\odot}$ RG star with an orbital period of $454.1\pm0.41\,\rm days$ (Brandi et al. 2009), and T CrB has a $\sim$$1.2\, M_{\odot}$ WD and a $\sim$$0.7\, M_{\odot}$ RG star with an orbital period of $\sim$$227.6\,\rm days$ (Kraft 1958; Belczy\'nski \& Miko\l{}ajewska 1998). Recently, Miko\l{}ajewska \& Shara (2017) suggested that the WD in RS Oph should be a CO WD by analyzing its spectra, which strongly supports that RS Oph will form an SN Ia. The binary parameters of RS Oph and T CrB are located in the regions for producing SNe Ia in Fig.\,3, which indicates that they can form SNe Ia in their future evolution.

There are some other alternatively pathways for producing SNe Ia via the symbiotic channel, e.g., the mass-stripping model, the aspherical stellar wind model, the tidally enhanced stellar wind model, etc.
(1) Hachisu et al. (1999) proposed a mass-stripping model to stabilize the mass-transfer process and avoid the formation of CE, in which the stellar wind from the WD collides with the RG surface and strips some of the mass from the RG. However, this process has not been identified by the observations.
(2) L\"u et al. (2009) assumed an aspherical stellar wind with an equatorial disk from a RG to investigate the symbiotic channel of SNe Ia, leading to a higher SN Ia rate. However, the results of L\"u et al. (2009) are strongly affected by the mass loss rate and the outflow velocity of the equatorial disk.
(3) Chen et al. (2011) adopted the tidally enhanced stellar wind assumption presented by Tout \& Eggleton (1988) to study the symbiotic channel, and found that the parameter space for producing SNe Ia will be enlarged. Nevertheless, Chen et al. (2011) also argued that the parameter space depends critically on the tidal wind enhancement parameter $B_{\rm w}$, which is still poorly understood.

In the symbiotic channel, some of matter would be blown away from the system before SN explosion, mainly originating from the thick optically wind process when $\dot{M}_{\rm 2}>\dot{M}_{\rm cr}$. These blown away matter would remain as circumstellar matter (CSM). We found that the CSM in SN Ia explosions have masses up to $\sim$$1.0\, M_{\odot}$. Some direct evidence have been found for existence of CSM in the normal SN Ia SN 2006X (e.g., Patat et al. 2007). The symbiotic channel may be a possible formation pathway for SN 2006X-like events (see also Patat et al. 2007; Voss \& Nelemans 2008).

The companions of the exploding WDs in the symbiotic channel will survive in SN Ia explosions. These companions can be identified by observations due to their anomalous properties (e.g., Wang \& Han 2010). Marietta et al. 2000 argued that about 96\%$-$98\% of the envelope in the RG stars will be stripped during SN Ia explosions, leading to the formation of single He WDs. We found that the surviving companion stars are single He WDs with masses in the range of $0.19$$-$$0.45\, M_{\odot}$. Some observations supported the existence of single low-mass He WDs with masses lower than $0.45\, M_{\odot}$ (e.g., Marsh et al. 1995; Kilic et al. 2007), which may correspond to the surviving companions of SNe Ia from the symbiotic channel (e.g., Wang \& Han 2010).

The mass-transfer prescription adopted here is similar to that presented in Kolb \& Ritter (1990). This prescription is based on a power-law adiabatic assumption, which is still under debate (e.g., Woods \& Ivanova 2011). We assumed that the mass-transfer occurs in the local sound velocity here. In fact, the mass-transfer may happen in the supersonic velocity when the radius of RG stars are larger than the Roche-lobe radius. In this case, our model might underestimate $\dot{M}_{\rm 2}$. However, the present work at least gives an upper limit for the SN Ia rate from the symbiotic channel.

In this Letter, we added an integrated mass-transfer prescription into the Eggleton stellar evolution code to investigate the symbiotic channel for producing SNe Ia. We evolved a large number of WD$+$RG systems, and obtained the SN Ia production space that is significantly enlarged. We then carried out a series of BPS computations with this parameter space, and found that the SN Ia rates are in the range of $\sim$0.2$-$$0.6\times10^{\rm -4}\,\rm yr^{\rm -1}$, which is 5 times larger than that from previous studies. The delay times of SNe Ia from the symbiotic channel range from $400\,\rm Myr$ to $11\,\rm Gyr$, contributing to SNe Ia in the intermediate and old populations. The surviving companions of SNe Ia from the symbiotic channel may evolve to $0.19$$-$$0.45\, M_{\odot}$ single He WDs. We suggest that the symbiotic channel is a possible way for the formation of SN 2006X-like events.
The rate of symbiotic novae could provides some constraints on the progenitor candidates and birthrates of SNe Ia from the symbiotic channel. The population of symbiotic novae is \textbf{still not} fairly well known, and GAIA will likely provide much better estimates on the distances and frequency.

\section*{Acknowledgments}
We thank Philipp Podsiadlowski, Xiangcun Meng and Hailiang Chen for their helpful discussions.
This work is supported by the 973 programme of China (No. 2014CB845700), CAS (Nos. QYZDB-SSW-SYS001 and KJZD-EW-M06-01), the NSFC (Nos. 11673059, 11673058, 11573016, 11521303, 11733008, 11322324 and 11390374) and Yunnan Province (Nos. 2013HA005, 2013HB097, 2014FB189 and 2017HC018).

\label{lastpage}

\begin{thebibliography}{}\label{thebibliography}
\bibitem[Belczy\'nsk \& Miko\l{}ajewska (1998)]{BM98} Belczy\'nski, K., \& Miko\l{}ajewska, J. 1998, MNRAS, 296, 77
\bibitem[Belczy\'nsk et al. (2000)]{Bel00}       Belczy\'nski, K., Miko\l{}ajewska, J., Munari, U., Ivison, R. J., \& Friedjung, M. 2000, A\&AS, 146, 407
\bibitem[Branch et al. (1993)]{BFN93}            Branch, D., Fisher, A., \& Nugent, P. 1993, AJ, 106, 2383
\bibitem[Brandi et al. (2009)]{Bra09}            Brandi, E., Quiroga, C., Miko\l{}ajewska, J., Ferrer, O. E., \& Garca, L. G. 2009, A\&A, 497, 815
\bibitem[Chen et al. (2011)]{Che11}              Chen, X., Han, Z., \& Tout, C. A. 2011, ApJ 735, L31
\bibitem[Chomiuk \& Povich (2011)]{Cho11}        Chomiuk, L., \& Povich, M. S. 2011, AJ, 142, 197
\bibitem[Chugai (2008)]{Chu08}                   Chugai, N. N. 2008, Astronomy Letters, 34, 389
\bibitem[Eggleton (1973)]{egg73}                 Eggleton, P. P. 1973, MNRAS, 163, 279
\bibitem[Eggleton et al. (1989)]{Egg89}          Eggleton, P. P., Fitchett, M. J., \& Tout, C. A. 1989, ApJ, 347, 998
\bibitem[Ge et al. (2010)]{Ge10}                 Ge, H., Hjellming, M. S., Webbingk, R. F., Chen, X., \& Han, Z. ApJ, 717, 724
\bibitem[Ge et al. (2015)]{Ge15}                 Ge, H., Webbink, R. F., Chen, X., \& Han, Z. 2015, ApJ, 812, 40
\bibitem[Graur \& Maoz (2013)]{Gra13}            Graur, O., \& Maoz, D. 2013, MNRAS, 430, 1746
\bibitem[Hachisu et al. (1996)]{hac96}           Hachisu, I., Kato, M., \& Nomoto, K. 1996, ApJ, 470, L97
\bibitem[Hachisu et al. (1999)]{hac99}           Hachisu, I., Kato, M., \& Nomoto, K. 1999, ApJ, 522, 487
\bibitem[Han et al. (2000)]{han00}               Han, Z., Tout, C. A., \& Eggleton, P. P. 2000, MNRAS, 319, 215
\bibitem[Han \& Podsiadlowski (2004)]{han04}     Han, Z., \& Podsiadlowski, Ph. 2004, MNRAS, 350, 1301
\bibitem[Howell (2011)]{how11}                   Howell, D. A. 2011, Nature Commun, 2, 350
\bibitem[Hoyel \& Fowler (1960)]{HF60}           Hoyel, F., \& Fowler, W. A. 1960, ApJ, 132, 565
\bibitem[Hurley et al. (2002)]{Hur02}            Hurley, J. R., Tout, C. A., \& Pols, O. R., 2002, MNRAS, 329, 897
\bibitem[Iben \& Tutukov (1984)]{IT84}           Iben, I., \& Tutukov, A. V. 1984, ApJS, 54, 335
\bibitem[Justham et al. (2009)]{Jus09}           Justham, S., Wolf, C., Podsiadlowski, Ph., \& Han, Z. 2009, A\&A, 493, 1081
\bibitem[Kato \& Hachisu (2004)]{kat04}          Kato, M., \& Hachisu, I. 2004, ApJ, 613, L129
\bibitem[Kenyon (1986)]{Ken86}                   Kenyon, S. J. 1986, The Symbiotic Stars. Cambridge Univ. Press, Cambridge
\bibitem[Kenyon et al. (1993)]{Ken93}            Kenyon, S. J., Mario, L., Miko\l{}ajewska, J., \& Tout, C. A. 1993, ApJL, 407, L81
\bibitem[Kilic et al. (2007)]{KSP07}             Kilic, M., Stanek, K. Z., \& Pinsonneault, M. H. 2007, ApJ, 671, 761
\bibitem[King et al. (2003)]{krs03}              King, A. R., Rolfe, D. J., \& Schenker, K. 2003, MNRAS, 341, L35
\bibitem[Kolb \& Ritter (1990)]{Kol90}           Kolb, U., \& Ritter, H. 1990, A\&A, 236, 385
\bibitem[Kraft (1958)]{Kra58}                    Kraft, R. P. 1958, ApJ, 127, 620
\bibitem[Kroupa (2001)]{Kro01}                   Kroupa, P. 2001, MNRAS, 322, 231
\bibitem[Li et al. (2015)]{Li15}                 Li, J., et al. 2015, RAA, 15, 1332
\bibitem[Li \& van den Heuvel (1997)]{li97}      Li, X., \& van den Heuvel, E. P. J. 1997, A\&A, 322, L9
\bibitem[Liu et al. (2018)]{Liu18}               Liu, D., Wang, B., \& Han, Z. 2018, MNRAS, 473, 5352
\bibitem[Liu et al. (2016)]{Liu16}               Liu, D., Wang, B., Podsiadlowski, Ph., \& Han, Z. 2016, MNRAS, 461, 3653L
\bibitem[Liu et al. (2017)]{Liu17}               Liu, D., Wang, B., Wu, C., \& Han Z. 2017, A\&A, 606, A136
\bibitem[Lubow \& Shu (1975)]{Lub75}             Lubow, S. H., \& Shu, F. H. 1975, ApJ, 198, 383
\bibitem[L\"u et al. (2006)]{Lv06}               L\"u, G., Yungelson, L., Han, Z. 2006, MNRAS, 372, 1389
\bibitem[L\"u et al. (2009)]{Lv09}               L\"u, G., Zhu, C., Wang, Z., \& Wang, N. 2009, MNRAS, 396, 1086
\bibitem[Maoz et al. (2010)]{mao10}              Maoz, D., Keren, S., \& Avishay, G.-Y. 2010, ApJ, 722, 1879
\bibitem[Maoz et al (2014)]{mao14}               Maoz, D., Mannucci, F., \& Nelemans, G. 2014, ARA\&A, 52, 107
\bibitem[Maoz et al. (2012)]{mao12}              Maoz, D., Mannucci, F., \& Timothy, D. Brandt 2012, MNRAS, 426, 3282
\bibitem[Marietta et al. (2000)]{Mar00}          Marietta, E., Burrows, A., \& Fryxell, B. 2000, ApJS, 128, 615
\bibitem[Marsh et al. (1995)]{Mar95}             Marsh, T. R., Dhillon, V. S., \& Duck, S. R. 1995, MNRAS, 275, 828
\bibitem[Meng et al. (2015)]{men15}              Meng, X., Gao, Y., \& Han, Z. 2015, IJMPD, 24, 14, 1530029
\bibitem[Miko\l{}ajewska (2012)]{Mik12}          Miko\l{}ajewska, J. 2012, Baltic Astronomy, 21, 5
\bibitem[Miko\l{}ajewska \& Shara (2017)]{MS17}  Miko\l{}ajewska, J., \& Shara, M. M. 2017, ApJ, 847, 99
\bibitem[Miszalski \& Miko\l{}ajewska (2014)]{mm14} Miszalski, B., \& Miko\l{}ajewska, J. 2014, MNRAS, 440, 1410
\bibitem[Munari \& Renzini (1992)]{Mun92}        Munari, U., \& Renzini, A. 1992, ApJL, 397, L87
\bibitem[Nomoto (1982)]{nom82}                   Nomoto, K. 1982, ApJ, 253, 798
\bibitem[Nomoto et al. (1984)]{Nom84}            Nomoto, K., Thielemann, F. K., \& Yokoi, K. 1984, ApJ, 286, 644
\bibitem[Patat et al. (2007)]{Pat07}             Patat, F., et al. 2007, Sci, 317, 924
\bibitem[Podsiadlowski et al. (2008)]{Pod08}     Podsiadlowski, Ph., Mazzali, P., Lesaffre, P., Han, Z., \& F\"orster, F. 2008, New Astro. Rev., 52, 381
\bibitem[Rodr\'iguez-Flores et al. (2014)]{Rod14} Rodr\'iguez-Flores, E. R., et al. 2014, A\&A, 567, A49
\bibitem[Totani et al. (2008)]{tot08}            Totani, T., Morokuma, T., Oda, T., Doi, M., \& Yasuda, N. 2008, PASJ, 60, 1327
\bibitem[Tout \& Eggleton (1988)]{TE88}          Tout, C. A., \& Eggleton, P. P. 1988, MNRAS, 231, 823
\bibitem[Voss \& Nelemans (2008)]{Vos08}         Voss, R., \& Nelemans, G. 2008, Nat, 451, 802
\bibitem[Wang \& Han (2010)]{WH10}               Wang, B., \& Han, Z. 2010, MNRAS, 404, L84
\bibitem[Wang \& Han (2012)]{wh12}               Wang, B., \& Han, Z. 2012, New Astron. Rev., 56, 122
\bibitem[Wang, Li \& Han (2010)]{WLH10}          Wang, B., Li, X.-D., \& Han, Z. 2010, MNRAS, 401, 2729
\bibitem[Webbink (1984)]{web84}                  Webbink, R. F. 1984, ApJ, 277, 355
\bibitem[Whelan \& Iben (1973)]{whe73}           Whelan, J., \& Iben, I. 1973, ApJ, 186, 1007
\bibitem[Willems \& Kolb (2004)]{Will04}         Willems, B., \& Kolb, U. 2004, A\&A, 419, 1057
\bibitem[Woods \& Ivanova (2011)]{Woo11}         Woods, T. E., \& Ivanova, N. 2011, ApJL, 739, L48
\bibitem[Yungelson \& Livio(1998)]{Yun98}        Yungelson, L., \& Livio, M. 1998, ApJ, 497, 168
\bibitem[Yungelson et al. (1995)]{Yun95}         Yungelson, L., \& Livio, M. Tutukov A., Kenyon S. J., 1995, ApJ, 447, 656
\end{thebibliography}
\end{document}